\documentstyle[12pt]{article}
\newcommand{\be}{\begin{equation}}
\newcommand{\ee}{\end{equation}}
\newcommand{\bea}{\begin{eqnarray}}
\newcommand{\eea}{\end{eqnarray}}
\newcommand{\nn}{\nonumber}
\begin{document}

\newcommand{\nd}[1]{/\hspace{-0.6em} #1}
\begin{titlepage}
\begin{flushright}
\end{flushright}
\begin{centering}
\vspace{.1in}
{\large {\bf Time-dependent perturbations \\
in two-dimensional String Black Holes  }} \\

\vspace{.4in}
{\bf G. A. Diamandis, B. C. Georgalas, X. Maintas} \\
\vspace{.05in}
University of Athens,
Nuclear and
Particle Physics
Section, Panepistimiopolis, Kouponia, 157 71 Athens,
Greece
\vspace{.05in}

and \\

\vspace{.05in}
{\bf N.E. Mavromatos}\\
\vspace{.05in}
Theory Division, CERN, CH-1211, Geneva 23, Switzerland  \\
\vspace{.3in}
{\bf Abstract} \\
\vspace{.05in}
\end{centering}
{\small
\paragraph{}
We discuss time-dependent perturbations (induced by matter fields)
of
a black-hole background in tree-level
two-dimensional string theory.
We analyse the linearized case and show the possibility
of having black-hole solutions with time-dependent horizons.
The latter
exist
only in the presence of time-dependent
`tachyon' matter fields,
which constitute the only propagating degrees of freedom in
two-dimensional string theory. For real tachyon field configurations
it is not possible to obtain solutions with horizons
shrinking to a point. On the other hand,
such a
possibility seems to be realized  in the
case of string black-hole models
formulated on higher world-sheet genera.
We connect this latter result
with black hole evaporation/decay at a quantum level.}
\par
\vspace{0.9in}
\begin{flushleft}
CERN-TH.6671/92 \\
UA-NPPS.9/92 \\
September  1992 \\
\end{flushleft}
\end{titlepage}
\paragraph{}
\newpage
\paragraph{}
Recently much attention has been concentrated
on
black-hole background configurations of
two-dimensional string theory
after Witten's observation \cite{witt} that
such models can be represented as exactly solvable
(finite) Wess-Zumino conformal field theories.
It was also suggested \cite{emn1,emn2} that such
theories will be consistent with quantum mechanics
even during the process of black-hole evaporation \cite{hawk}.
The reason for this
is the existence of
a particular type of infinite-dimensional
quantum hair, the so-called
`W-hair', carried by
the black holes
as a result of an enormous stringy gauge symmetry
that mixes the various string levels
\footnote{In the case of
two-dimensional target
space the only propagating string
degrees of freedom are the massless `tachyon'
fields. The higher-spin
string states
are
non-propagating
with definite energies and momenta \cite{gro,pol};
nevertheless, these (quasi-topological) modes
play an important r\^ ole in the physics \cite{emn1,emn2}.}.
Due to these symmetries, the two-dimensional phase space
of the
matter tachyon fields is preserved under time evolution,
provided that one takes into account the
discrete massive string states.
As
a result
the modifications \cite{ehns}
to the usual quantum mechanical evolution
equations
for the density matrix of the matter
system
interacting with the black hole are viewed as artefacts
of the truncation of the string spectrum to the
massless modes only \cite{emnqm}.
For instance, it has
been shown in \cite{chaudh} that the exactly
marginal operator that turns on static (massless) tachyon
backgrounds in the coset black-hole
model of \cite{witt},
{\it necessarily}
involves
the entire
spectrum of
{\it massive} string
modes.
This makes the system of the light modes
of the string in the presence of black-hole backgrounds
`open', thereby leading to apparent
modifications of quantum mechanics for the massless
string states \cite{emnqm} in a string field theory framework.
\paragraph{}
As argued in \cite{emn3}
the stringy black holes can `evaporate', but
unlike the local field theory case \cite{hawk}
their
evaporation
is non-thermal,
resembling an ordinary decay of a massive string state.
There are various decay channels and selection rules
that characterize the decay process, which involve
higher excited string (discrete) states, in addition
to the massless tachyon fields.
Their precise form is
not yet fully known, although
significant advances have been made in this direction
\cite{emn5,kle}.
\paragraph{}
Motivated by these results,
we would like to examine in some detail,
in the present letter,
time-dependent perturbations of
static black-hole solutions of
two-dimensional (target) space-time
string theory. We should remark that
to our knowledge time-dependent perturbations
of local field theory black holes induced
by scalar fields have been studied, so far,
explicitly
only in the context of five-dimensional
Kaluza-Klein theories,
where the scalar field is associated with the
component of the metric pertaining to the extra-dimension
\cite{tom}. In a similar context it has also been argued
\cite{venez}
that $O(d,d)$-transformations of static two-dimensional
string
black-hole
backgrounds could absorb the singularity, at the cost of
introducing an
extra target-space
dimension. Our approach here
will be different
in that we shall examine time-dependent
perturbations of two-dimensional
string
black holes by
avoiding the introduction of extra dimensions.
We shall show that
there are
solutions,
at least in
the linearized approximation,
describing time-varying horizons,
but only in the presence of time-dependent
configurations for the tachyon fields.
However it seems that for real matter-field
configurations it is not possible to obtain
solutions with horizons shrinking down to a point,
a possibility that seems to be realized in the
case of complex matter fields. As we shall discuss,
this latter result might be connected with the
decay of the black hole induced by higher genera \cite{emn3}.
As is well-known in string theory, higher
genus effects can be effectively
represented as additional
renormalization counterterms
of the
tree-level $\sigma$-model couplings \cite{susk}.
The effect of higher genera
is to {\it add} extra
{\it marginal} operators which do not exist
in the
tree-level $\sigma$-model background theory.
Slightly relevant
deformations lead to imaginary parts in the effective
target-space action, as a result of the circulation
of these modes along the string loops.
\paragraph{}
We start our discussion
from the low-energy effective action
with dilaton, tachyon and graviton fields
for a two-dimensional string theory. For simplicity
we restrict ourselves to one ($\sigma$-model) loop order,
which from an effective field theory point of view
corresponds to a large-$k$ Wess-Zumino model, with $k$ the
level parameter \cite{witt}. In this truncated theory
conformal invariance conditions should
always be understood as
{\it approximate} solutions where corrections
of the order of the Planck mass are suppressed.
The action takes the form
\be
I_{eff}=\frac{1}{2g^2}\int d^2 x \sqrt{G} e^{\Phi}
\{ R + (\partial \Phi)^2 + (\partial T)^2 -8T^2 + \Lambda \}
\label{eff}
\ee

\noindent where $G_{\mu\nu}$ is a (Euclidean signature) metric
in a two-dimensional target
space-time, $g$ is the gravitational coupling
and $\Lambda$ is a cosmological
constant arising from the non-criticality of the dimension
of the string space-time \cite{aben}.
The usual ambiguities of the tachyon potential in string
theory \cite{banks} have been {\it fixed} here to
quadratic configurations for the tachyon field.
The equations of motion obtained from (\ref{eff}) are,
\bea
\nn D_{\mu}D_{\nu}\Phi &=& G_{\mu\nu}\{ D^2 \Phi + \frac{1}{2}(\partial
\Phi)^2 - \frac{1}{2}(\partial T)^2 + 4T^2 - \frac{1}{2}\Lambda \}
+ \partial_{\mu}T \partial_{\nu}T \\
\nn 0 &=& -2D^2 \Phi - (\partial \Phi)^2 + R + \Lambda - 8 T^2 +
(\partial T)^2 \\
0 &=& D^2 T + (\partial _\mu
\Phi \partial ^\mu T) + 8T .
\label{eqs}
\eea

\noindent In the  absence of matter, $T=0$, the first
of these equations imposes the solution \cite{witt}
$\Phi=ln ch^2 Q r $ and the invariant line
element acquires the form
$ ds^2 =dr^2 + tgh^2 Q r dt^2 $, where $Q^2=\Lambda$.
The existence of such a solution is guaranteed
by the fact that when $T=0$ the vector
$\partial_{\mu}\Phi$ is {\it hypersurface-orthogonal}
and {\it geodesic}, according to a
theorem
of general relativity \cite{stef}. Changing variables \cite{wad},
$ln ch^2 Qr =2Q\rho - ln\alpha$, leads to the gauge
$\Phi=2Q\rho -ln\alpha$
($\alpha$ is an arbitrary constant),
in which the line element acquires the form
\be
ds^2=\frac{d\rho ^2}{1-\alpha e^{-2Q\rho}} + (1-\alpha e^{-2Q\rho})dt^2 ,
\label{bh}
\ee

\noindent where $\alpha$ plays the role of the mass of the black hole
\cite{witt,wad}. It is interesting to note that in the limiting
case $\alpha \rightarrow 0$, one discovers
the $Q$-graviton
discrete state of Polyakov \cite{pol}.
The excitation of this, as well as
higher-spin topological string states,
might thus
be interpreted as a general feature of the
last stage of the black-hole evaporation \cite{emn1},
and has important consequences for the existence of
the $W$-hair of the black hole, and the consistency
with quantum mechanics \cite{emn2}.
\paragraph{}
To incorporate time dependence in these solutions, in the
presence of non-trivial matter, $T \ne 0$, it is
convenient to write them in the form, following \cite{deal} :
\bea
\nn  D_{\mu} D_{\nu}\Phi -
\frac{1}{2}G_{\mu\nu}D^2 \Phi &=&
\partial_{\mu}T \partial_{\nu}T - \frac{1}{2}G_{\mu\nu}T
(\partial T)^2 \\
\nn R &=& D^2 \Phi - (\partial T)^2     \\
\nn D^2 \Phi + (\partial \Phi)^2 &=& \Lambda - 8T^2 \\
0 &=& D^2 T + (\partial_{\mu} \Phi \partial^{\mu} T)
+ 8T .
\label{daw}
\eea

\noindent From the
five independent equations (\ref{daw})
we may relax one, since
we are effectively
working with conformal invariance conditions
of $\sigma$-models and hence the
Curci-Paffuti relation \cite{CP}
applies
\be
\frac{1}{2}D_{\nu}\beta^{\Phi}=(D^{\mu} + D^{\mu}\Phi)
\beta_{\mu\nu}^{G} +  \frac{1}{2}(D_{\nu} T)\beta^T .
\label{cur}
\ee

\noindent Below
we choose to relax the second of the equations (\ref{daw}).
Using the {\it hypersurface orthogonality} property of the
gradient vector $\partial_{\nu} \Phi$ \cite{stef}, we
are free to choose the gauge
\bea
\nn  \Phi &=& Q\rho \\
ds^2 &=& \frac{1}{g(\rho,t)}d\rho^2 + A(\rho,t)g(\rho,t)dt^2 ,
\label{gaug}
\eea

\noindent where $g,A \ge 0$ outside the (outer) horizon,
and $A$ is a {\it regular} function of $\rho, t$.
Using the notation  $\partial _{t}$   for a time derivative and $'$
for a spatial-one, we can write the equations (\ref{daw})
in component form
\bea
\nn \frac{\partial _{t} g}{g} &=&  \frac{2}{Q}\partial _{t} T T'  \\
\nn \frac{A'}{A} &=& -\frac{2}{Q} (T')^2 +
\frac{2}{Ag^2}\frac{(\partial _{t} T)^2}{Q}     \\
\nn g' + (g-1)Q &=& -\frac{g}{2}
\frac{A'}{A} - \frac{8T^2}{Q} \\
\frac{\partial _{t} g} {Ag^2} \partial _{t} T -
\frac{g}{2} \frac{A'}{A} T' &=&
gT'' + (g' + gQ)T' + 8T + \frac{1}{Ag}\partial _{t}^2 T
- \frac{1}{2g} \frac{\partial _{t} A}{A^2} \partial _{t} T  .
\label {comp}
\eea

\noindent It is readily checked that
the above system of four
equations with three unknowns
is compatible.
Following the standard {\it iterative} method
to solve this non-linear system, we introduce a
{\it small} parameter $\epsilon$, which serves as
a book-keeping parameter of the order of linearity.
We
rewrite
the
system of equations (\ref{comp})
as
\bea
\nn \frac{\partial _{t} g}{g} &=&
 \frac{2\epsilon}{Q}\partial _{t} T T' \\
\nn g' + (g-1)Q &=& \frac{g\epsilon}{Q} (T')^2 -
\frac{\epsilon (\partial _{t} T)^2 }{QgA} - \frac{\epsilon 8 T^2}{Q} \\
\nn g T'' + (g' + g Q)T' + 8T +
\frac{1}{g\sqrt{A}}
\partial _{t}
(\frac{ \partial _{t} T}{\sqrt{A}}) &=&
\frac{\epsilon g}{Q}(T ')^3 +
\frac{\epsilon}{Qg}(\frac{\partial _{t} T}{\sqrt{A}})^2 T'  \\
\frac{A'}{A} &=&-\frac{2\epsilon}{Q}(T')^2 +
\frac{2\epsilon}{Qg^2A}(\partial _{t} T)^2
\label{pert}
\eea

\noindent and look for solutions for $g$
and $A$ of the form $g=g_0 + \epsilon g_1 + ... $,
$ A= A_0 +\epsilon A_1 + ...$  In the static case, the system of
equations  (\ref{pert}) is
satisfied
with the choice $g_0 =A_0 =1$  and
$T_0 =(\mu_1 + \mu_2 \rho)e^{-\frac{1}{2}Q\rho}$. This corresponds
to the solution of \cite{deal}, and it is compatible with the
result for the configurations of the tachyon field expected
on general grounds from Liouville theory \cite{polch}, where
$\mu _{1}$ might be identified with the Liouville (world-sheet)
cosmological constant.
\paragraph{}
The non-static case can be dealt with {\it iteratively}.
As an {\it initial step}
we choose
\be
\nn  g_0 = 1 \qquad , \qquad A_0 = f(t)^2
\label{cond}
\ee

\noindent with $T_0$ satisfying the linear equation
\be
          T_0'' + QT_0' + 8T_0 + \frac{1}{f}
\partial _{t}(\frac{\partial _{t} T_0}{f}) = 0 .
          \\
\label{step}
\ee

\noindent In this and the following expressions we only
consider the time-dependent part of the matter (tachyon)
fields, because we are only
interested in the
effects of time-dependent perturbations on the
black-hole background.
\paragraph{}
To first order in $\epsilon$ the equations
for $A_1, g_1$ become:
\bea
\nn \partial _{t} g_1 &=& \frac{2}{Q}\partial _{t} T_0 T_0 ' \\
g_1 ' + g_1 Q &=& \frac{1}{Q} T_0 '
- \frac{1}{Q}\frac{(\partial _{t} T_0)^2}{f^2}
- \frac{8}{Q}T_0^2  \\
\nn A_1 '&=& \frac{2}{Q} (-f^2 (T_0 ')^2 +(\partial _{t} T_0)^2 ) .
\label{first}
\eea

\noindent Making use of the transformation $g_1 = e^{-Q\rho}u_1 $
we can write the equations for $g_1$  in the form
\bea
\nn \partial _{t} u_1 &=& \frac{2}{Q}e^{Q\rho} \partial _{t}T_0 T_0' \\
u_1 '&=&\frac{e^{Q\rho}}{Q}(T_0'^2 -
(\frac{ \partial _{t}T_0}{f})^2 - 8T_0^2 ) .
\label{ueq}
\eea

\noindent Setting $\chi = \int_0^t f(\tau) d\tau$
and $T_0=e^{-Q\rho}{\hat T}_0$, we observe that
${\hat T}_0$ is harmonic in $(\rho,\chi)$-space,

\be
    (\partial_{\rho\rho} + \partial_{\chi\chi}){\hat T}_0 = 0 .
\label{harm}
\ee

\noindent The general solution is ${\hat T}_0 = Re F(z)$,
where $F(z)$ is an arbitrary analytic function of $z=\rho + i \chi (t)$.
It is not difficult to see, then, that the general solution
for $g(\rho, \chi (t))$ takes the form
\bea
\nn g(\rho,\chi)=1+\frac{1}{Q}
\epsilon e^{-Q\rho} \{&2&
 \int_{\chi_0}^{\chi} d\lambda
[(\partial_{\chi} {\hat T}_0)
(\partial_{\rho} {\hat T}_0)   - \frac{Q}{4}
(\partial_\chi {\hat T}_0)^2 ] |_{\rho =0,\chi =\lambda} + \\
&+& \int_{\rho_0}^{\rho} d\lambda [(\partial_\rho {\hat T}_0)^2
-(\partial_\chi {\hat T}_0)^2 -\frac{Q}{2}
(\partial _{\rho} {\hat T}_0)^2 ]|_{\rho =\lambda} \} .
\label{final}
\eea

\paragraph{}
The general solution for $A(\rho, t)$ is
given by
\be
A(\rho, t)=f(t)^2 \{ 1+\epsilon \int _{\rho_0}^{\rho}
 e^{-Q\lambda}[(\partial_{\chi} {\hat T}_0 )^2 -(\partial_{\rho}
{\hat T}_0 - \frac{Q}{2} {\hat T}_0)^2 ]|_{\rho = \lambda}d\lambda \}
+ h(t) .
\label{alph}
\ee

\noindent where $f(t) \ge 0$, $h(t)$ are functions (otherwise
arbitrary) that
guarantee the regularity of $A(\rho, t)$. Without loss of generality
we
set $f(t) = 1$ for simplicity.
\paragraph{}
We can now arrive at some general conclusions
regarding the solutions.
We assume a polynomial form for the tachyon, which is
compatible with a weak-field expansion:
$F(z)=\alpha_0 + \alpha_1 z + ... \alpha_n z^n $,
with $\alpha_i$, $i=1,2,...n$ real. In
case
one is
interested only
in the time-dependent parts of the tachyon, as most
relevant for the question on the evaporation
addressed in this work,
one can drop the
static part \cite{deal} and hence set $\alpha _0=\alpha _1=0$.
\paragraph{}
When $n=2m+1$, $m=$ positive integer, we observe that
\bea
 \nonumber
 g(\rho ,\chi (t)) &=& 1 + e^{-Q\rho} (c +
\alpha _{2m}^2 \chi ^{4m}\{ \frac{2m+1}{Q} \kappa - \frac{1}{2}
+ [\frac{(2m+1)^2}{Q}\kappa ^2 - \\
&-& (2m+1)\kappa]\rho -
\frac{(2m+1)^2}{2}  \kappa ^2   \rho ^2 \} + O[  \chi ^{4m-2} ]) ,
\label{even}
\eea

\noindent with $\kappa \equiv \alpha_{2m+1}/\alpha _{2m} $.

\paragraph{}
The appearance of the negative coefficient of the
$\rho ^2$ term
{\it always }
implies
the existence of an {\it event
horizon} which is slightly expanding
with increasing time (see fig. 1a).
\paragraph{}
The existence of non-vanishing horizons also characterizes
the {\it even} order ($n=2m$)
polynomial solutions. This becomes clear
already
from the boundary value $g(0,\chi (t))$,
\be
g(0,\chi (t) )=1 + c - \frac{\epsilon}{2}\alpha _{2m}^2 \chi (t) ^{4m}
+ O[ \chi ^{4m-2} ] .
\label{oddp}
\ee

\noindent The difference
in this case, as compared with the previous one,
is that the horizon increases considerably
with increasing time (see fig. 1b).
This class of solutions
might be
considered as corresponding to {\it absorption} by the
black hole of time-dependent matter; in view of energy
conservation of the matter-black-hole system, such a process
will increase the mass of the black hole and consequently
its horizon. In  principle the increase is unbounded.
However, from a conformal
field theory point of view,
the scattering of light particles off a static black-hole background
\cite{verl,emn5} leads to excitations of discrete (higher-spin) states
that constitute the internal degrees of freedom of the black hole.
The latter then decay, emitting light particles whose number
is restricted by appropriate selection rules \cite{emn5}.
An important feature is
the
irreversibility of the
decay process
that seems to
characterize the
stringy black holes \cite{emndua}.
In
view of these results,
solutions with horizons expanding to spatial infinity
should be considered as
{\it not representing} an exact conformal field theory.
Probably when the rest of the backgrounds,
corresponding to the topological modes of the string,
are taken into account,
the phenomenon of the
uncontrollable increase in the horizon
disappears.
\paragraph{}
We expect
the above conclusions,
concerning the existence of
event horizons non-shrinkable to a point,
to hold
for any solution of (\ref{harm})
for the tachyon field. A simple argument
for this is based on
the fact that
one is mainly
interested in the behaviour of the solution
at finite time and space intervals, and we know that
any analytic function in the region $0 \le \rho \le L$,
$0 \le \chi \le T$, with $ T    >> 0 $, can be resolved in
orthogonal polynomials whose highest power coefficient
is real.
\paragraph{}
It should be noticed at this stage that a similar behaviour
of non-shrinking horizons
characterizes the second-order (in $\epsilon $)
solutions of the
non-linear system (\ref{pert}), as becomes clear from the relevant
expressions for the metric field.
For the case of $  (2m + 1)$-degree
polynomials, the second-order analysis yields
for the metric
at the origin $\rho =0 $
\be
g_{2}(0,\chi ) = \epsilon ^2 \alpha ^4 \chi ^{8m} ( \frac{1}{8}
+ \frac{1}{4}\lambda + \frac{3}{4} \lambda ^2 - \frac{1}{2}
\lambda ^3  - \frac{3}{2} \lambda ^4 ) ,
\label{secord}
\ee

\noindent with $\lambda \equiv n\kappa /Q$, and $n$ denotes
the degree of the
solution.
\paragraph{}
The situation is slightly different for the even-degree polynomials
$n=2m$, where the second-order correction in the
leading-$\chi $ behaviour
becomes positive:
\be
g_2 (\rho , \chi ) = 1 +  e^{-Q\rho} [ c - \frac{\epsilon}{2}
\alpha ^2 \chi ^{4m} + \frac{1}{8} \epsilon ^2 \alpha ^4
\chi ^{8m} e^{-Q\rho} + \dots ] .
\label{secorddio}
\ee

\noindent However, as becomes clear from fig. 3, this still
leads to non-shrinking horizons for the black-hole solution.
\paragraph{}
The above considerations
are indicative of some sort of general
behaviour that characterizes tree-level stringy black holes,
which appear as stable solutions of the conformal invariance
conditions (in the approximation where
the Regge slope $\alpha '\rightarrow0 $). Higher-genus
corrections on the world-sheet,
however, can change the situation drastically,
as we shall discuss
shortly below.
\paragraph{}
At the moment,
some comments are in order concerning the perturbative
nature of the polynomial configurations for the matter fields.
The solutions (\ref{even})--(\ref{secorddio}) have been obtained
for $\epsilon $ small. This, however, does not prevent one
from analysing the behaviour
at large
times,
provided that $\epsilon \alpha ^2 \chi^{4m} \le 1$, and thus
justifying restriction
to the highest
relevant powers. It should be remarked at this stage
that even in the region $\epsilon \alpha \chi ^{4m} > 1 $,
where perturbative calculations break down, the form of the
solutions, at least up to second order in $\epsilon $,
remains the same, as can be seen by simple
inspection of the relevant equations. However this is only a formal
observation, and
in order
to extrapolate the results for $\chi \rightarrow \infty$ safely,
one needs non-perturbative
information.
\paragraph{}
The above solutions have been obtained in Euclidean formalism,
but can be analytically continued to Minkowski space-times.
This does not affect our conclusions,
given that the
time dependence of the relevant terms
appears through powers of $\chi$ that are
multiples of $4$. It should be noticed, though, that
at the level of
the Wess-Zumino representation of the two-dimensional
(static) black-hole solution the spectrum of the
Euclidean black hole is not the same as that of the
Minkowskian one \cite{distl}.
Despite this, there are certain features of the
Euclidean formalism,
as for example the
formal character of selection rules in black-hole
decay \cite{emn5}, that
can be transcribed to the Minkowski (physical)
case
by simple analytic continuation.
It is in this sense that we apply this method here.
{}From a conformal field theory point of view, we are implicitly
working with Minkowski space coset models (and their
deformations to include matter fields),
and analytically
continue to Euclidean signature
only
at the level of the effective action
for computational
easiness.
\paragraph{}
The above considerations have indicated
that there
exist no
time-dependent
solutions with
horizons shrinking to a point
in the
case of
two-dimensional
stringy black holes formulated at a
tree-level
on the world sheet.
In fact, it becomes evident
from (\ref{even})
that
it is not
possible to obtain a shrinking horizon
with a real tachyon field, at least within a
perturbative framework.
On the other hand, such a possibility may
arise
in the case of
complex tachyon fields. The latter may be
considered as an effective description of
higher-genus world-sheet effects.
Indeed, as shown in \cite{emn3} as a result of the
regularization of
modular infinities by
analytic continuation,
the one-string-loop (torus)
analysis shows the appearance of a phase in front of the
Einstein-dilaton terms in the effective action (\ref{eff}), but
{\it not} in front of the matter (tachyon) sector.
In order to keep the space-time geometry real,
this would lead, at least na\" ively, to
an effectively complex tachyon
field, which would invalidate
the analysis leading to (\ref{even}). In this way one could
obtain
black-hole solutions with
horizons shrinkable to a point, thereby allowing for a
possibility of `evaporation'/decay along the
lines
of  ref. \cite{emn3}.
However a rigorous proof
still awaits a
consistent genus expansion of string theory
in a closed form. For the two-dimensional (target) space-time
case, this could be achieved in
a matrix-model
representation of the stringy black hole.
{}From this point of view,
the situation would be described by induced
time-dependent
deformations in the light sector of string theory.
\paragraph{}
The evaporating
black-hole solutions in string theory would be
characterized by the coherence-preserving
target space $W_{\infty}$-symmetries (W-hair)
which
are responsible for a mixing of the
string mass levels \cite{emn1,emn2,emn5}. The size of the
black-hole could be macroscopic or
microscopic. Microscopic (virtual) black holes
have been argued \cite{hawk} to appear in fluctuations
of the metric field in a quantum gravity framework.
In string field theory,
such virtual black holes would lead to
the apparent modifications of the light string mode quantum
mechanics as a result of the level-mixing
$W_{\infty}$-symmetries
\cite{emnqm}.
Although at
present we have not constructed explicitly such exact
string
configurations, however,
we expect them to exist
on the basis of the above discussion and quite
general arguments \cite{emn3}.
We leave these considerations, as well as attempts to solve
exactly the systems of coupled equations discussed above, for
future work.

\paragraph{}
\noindent {\Large{\bf Acknowledgements}} \\
\par We are grateful to A. Petridis for his invaluable
help with the computer. One of us (G.A.D.) thanks the Theory
Division of CERN for its hospitality during the last stage
of this work.

\newpage
\noindent {\Large {\bf Figure Captions}}

\paragraph{}

\noindent {\bf Fig. 1}.  The behaviour of the metric tensor
as a function of $\rho$ and $\chi$
to order $\epsilon$ in the case of : (a) odd-degree polynomial
solutions, and (b) even-degree polynomial solutions.
Clearly one sees the existence of black-hole
horizons non-shrinkable
to a point.

\paragraph{}

\paragraph{}

\noindent {\bf Fig. 2}.  The behaviour of the metric tensor
as a function of $\rho$ and $\chi$
to order $\epsilon^2
$ in the case of : (a) odd-degree polynomial
solutions, and (b) even-degree polynomial solutions.
The existence of horizons non-shrinkable to a point
also characterizes the solutions to this order.

\end{document}